\documentclass[prb,twocolumn,showpacs,amsmath,amssymb]{revtex4}

\usepackage{graphicx}
\usepackage{dcolumn}
\usepackage{bm}
\usepackage{revsymb}
\usepackage{txfonts}

\begin{document}

\title{Edge state on hydrogen-terminated graphite edges investigated by scanning tunneling microscopy}

\author{Yousuke Kobayashi}
\email{ykobaya@chem.titech.ac.jp}
\author{Ken-ichi Fukui}
\author{Toshiaki Enoki}
\affiliation{Department of Chemistry, Tokyo Institute of Technology, 2-12-1, Ookayama, Meguro-ku, Tokyo 152-8551, Japan}

\author{Koichi Kusakabe}
\affiliation{Graduate School of Engineering Science, Osaka University, 1-3, Machikaneyama-cho, Toyonaka, Osaka 560-8531, Japan}

\date{\today}

\begin{abstract}
The edge states that emerge at hydrogen-terminated zigzag edges embedded in dominant armchair edges of graphite are carefully investigated by ultrahigh-vacuum 
scanning tunneling microscopy (STM) measurements. The edge states at the zigzag edges have different spatial distributions dependent on the $\alpha$- or $\beta$-site edge 
carbon atoms. In the case that the defects consist of a short zigzag (or a short Klein) edge, the edge state is present also near the defects. The amplitude of the edge state 
distributing around the defects in an armchair edge often has a prominent hump in a direction determined by detailed local atomic structure of the edge. The tight 
binding calculation based on the atomic arrangements observed by STM reproduces the observed spatial distributions of the local density of states.
\end{abstract}

\maketitle

\section{\label{sec:level1}Introduction}
Low dimensional nanosized carbon materials include carbon nanotubes and a nano-sized graphene sheet (nanographene), where nanographene is two-dimensional 
(2D) flat hexagon network with open edges and nanotube has 1D structure made by rolling up a graphene sheet. These carbon materials have peculiar $\pi$-electron 
systems because of their characteristic structures, according to experimental and theoretical reports in this decade \cite{enoki1,enoki2,ando1,tomanek1}. 
The peculiarity of the electronic structures of these carbon materials originates from the dimensionality, the size, the chirality, and the presence of the edges. 
Carbon nanotubes are characterized by the tube diameter and the chirality \cite{ando2}, and the electronic structure can be modified also by the presence of ends 
if the length of the tube is small \cite{nanotube}. In this sense, the electronic structure of nanographene is characterized more by the presence of ends that encircle 
the finite-sized 2D sheet. Indeed, related to the electronic structure of the ends of carbon nanomaterials, edge state, which originates from the local electronic structure 
at the edge of graphene, has been quoted as an explanation of experimental findings in the unconventional electronic and magnetic properties of nanographite/graphite 
\cite{shibaya,klusek}. Recently ferromagnetism has been reported for defects of graphite, where the local spin arrangement of carbon atoms near 
defects and graphite edges seems to be responsible for the ferromagnetism \cite{ferroex1,ferroex2,ferroth,esquinaz}. The electronic state at the defects can be related 
to the electronic structure around the edges of graphite or carbon nanotubes.

The electronic state of the edges is dependent on the local geometric structure, for example, zigzag and armchair edges. An edge state appears as a non-bonding 
$\pi$-electron state that originates from the 2$p_{z}$ orbitals of carbon atoms near a zigzag edge, according to theoretical studies \cite{kobayash,fujita,nakada}. 
The edge state of a zigzag edge is featured by a flat band near the Fermi energy, and the distribution of the edge state is dependent on a combination of zigzag 
and armchair edges. Theoretical reports suggests that a Klein edge, which is defined as a zigzag edge with each edge carbon atom being bonded to an additional 
carbon atom, also has the edge state similar to that of a zigzag edge \cite{klein1,klein2}. Interestingly, ferromagnetism is theoretically predicted for nanographene 
ribbon with a zigzag edge at one side and a kind of a Klein edge at another side \cite{kusakabe}.

For verifying the presence of the edge state, some experimental efforts have been made recently on the basis of scanning tunneling microscopy and spectroscopy (STM/STS) 
\cite{edge, niimi}. However, their experimental condition is not suitable, that is, observations of graphite edge were performed in the ambient atmosphere. 
Around the edges prepared in the ambient condition, functional groups terminate the edge carbon atoms. These species strongly modify the electronic structure of the edges. 
In addition, weakly adsorbed species alter the electronic features due to chage transfer. In theoretical works in Refs.~\onlinecite{kobayash} and ~\onlinecite{miyamoto}, 
the edge state come from the hydrogen-terminated zigzag edge. Accordingly, hydrogen-termination of graphite edge after heat-treatment under an ultrahigh-vacuum (UHV) condition is essential 
for observing the edge state. Besides, this enables us to control the electronic structure and the Fermi energy and 
to distinguish the edge state near the Fermi energy from the $\sigma$-dangling bond state by terminating dangling bonds with hydrogen atoms at the graphite edge. 
Recently, the present authors reported STM/STS observations of graphite samples with the well-defined edges that were prepared by hydrogen termination under 
an UHV condition \cite{ykobaya}. This study clearly confirmed the presence of the edge state and revealed its peculiar features for the first time. 
Actually the edge state, which lies near the Fermi energy, spatially distributes near a zigzag edge and a partial zigzag edge that is embedded in an armchair edge. 
Interestingly, the local electronic state that causes the edge state distributes in a directive array depending on the local geometric structure of graphite edges.

On the basis of our previous experimental results, successive observations of the edge state in a local space is favored to accumulate comprehensive knowledge about the edge state 
related to the atomic arrangement at edges. In the present paper, we report STM investigations of local electronic structures of hydrogen-terminated partial-zigzag edges 
and their dependence on the local geometric structure to clarify the distribution of the edge state comprehensively, as a succession of the previous report in Ref.~\onlinecite{ykobaya}. 
It is further confirmed that the edge states are very sensitive to the detailed geometric structure of graphite edges.

\section{\label{sec:level2}Experimental}

All STM images were taken in constant-height mode at $V_{S}=0.02$ V and $I=0.7$ nA, using Pt-Ir tips by a UHV-STM (Unisoku Co.). Brightness, which is magnitude of tunneling 
current, of the STM image at this tunneling condition is qualitatively proportional to local density of states (LDOS) near the Fermi energy. The sample is step edges of highly oriented 
pyrolytic graphite (HOPG) \cite{affoune}. It was heat-treated at 800 $^{\circ}$C under UHV to eliminate functional groups including oxygen atoms and was exposed to atomic hydrogen 
at $1\times 10^{-6}$ Torr for hydrogenation of the edges, just after the heat treatment \cite{co}. The condition for the hydrogenation of the edges was the same as that for hydrogenation 
of the Si(100) surface to make a monohydride surface \cite{boland}. After hydrogen termination of graphite edges of the sample, the UHV condition was kept until the end 
of STM measurements. Adsorbed contaminants on the edges and graphite surface can be removed by reaction with pure hydrogen during the hydrogenation process. 
By several repeats of the heat treatments and hydrogenation in the preparation chamber, well-ordered hydrogen-terminated edges were created \cite{zecho1,zecho2}.

The 2D LDOS mappings were calculated using the tight-binding approximation for $AB$-stacked double-layer graphene. The first layer represents 
the top graphene layer with edges and the second layer represents the graphite substrate. Only 2D LDOS mapping of the first layer is shown in figures. The relative value of 
the LDOS was accumulated in the range of 0.05 eV near the Fermi energy near the Fermi energy. The area of the circle on each lattice point in figures of calculated results is 
proportional to the LDOS. The resonance integral and the overlap integral were parameterized using the Slater-Koster parameters \cite{slater} and were determined for 
the 2$s$ and 2$p$ orbitals of carbon and the 1$s$ orbital of hydrogen. The structural dependence of the parameters was determined following the previous literature for 
carbon \cite{papacon}. For carbon-hydrogen bonding, we fitted the parameters of hydrogen to reproduce the band structure of graphene strips with zigzag edges obtained 
by a first-principles calculation with the local density approximation \cite{perdew,yamauchi}. Several percentages of displacements of carbon atoms near each edge were 
neglected in the H\"{u}ckel approximation. This makes the calculation tractable without harming essential features in the DOS.

\section{\label{sec:level3}Results}

Clear STM images of edges were observed at the steps of HOPG samples after the heat-treatment under UHV followed by an exposure to atomic hydrogen to terminate the 
step edges. It was quite in contrast with the observation without such treatments, where various functional groups were randomly bonded to the edges.

\begin{figure}[here]
\includegraphics[width=8.5cm, clip]{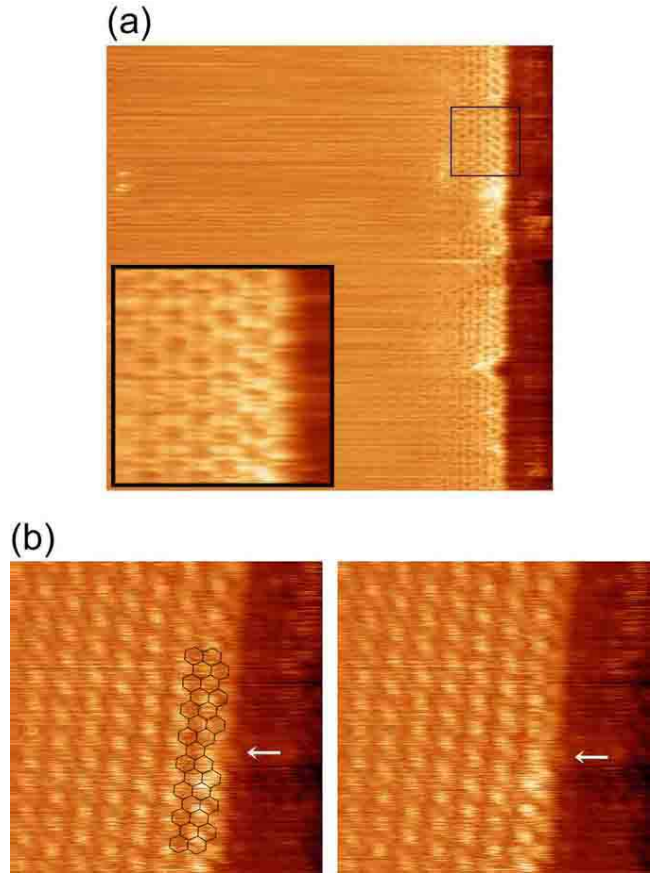}
\caption{\label{fig:edge1}(Color online) (a) STM image (20.5 $\times$ 20.5 nm$^{2}$) of an armchair edge with defect points. Inset is a magnified image (3.1 $\times$ 3.1 nm$^{2}$) of 
the square region on the right top. (b) STM image (3.9 $\times$ 3.9 nm$^{2}$) of an armchair edge with a dent (indicated by an arrow) that is made by removal of two 
carbon atoms from the homogeneous armchair edge. The right panel is the same image to that at the left panel without drawing the honeycomb lattice.}
\end{figure}

Figure 1(a) shows an STM image of a hydrogen-terminated armchair edge at the step of the HOPG sample. The length of the armchair edge was much larger than the scanned size. 
STM observations revealed that the length of the armchair edge tended to be extended about a few hundreds of nanometers. However, such armchair edges do not necessarily 
have a homogeneous structure, but have defects such as protrusions or dents. Among these defects, two-atoms dent was observed as shown in Figs. 1(b), as the smallest 
defects in the armchair edge. Larger LDOS, which was characterized by brighter contrast in the STM image, was observed in the vicinity of the lower side of the dent. In the 
observations of the armchair edge, other types of defects, that is, those created by carbon atom rows attached to the armchair edge, were also observed. These will be shown later.

\begin{figure}[here]
\includegraphics[width=8.5cm, clip]{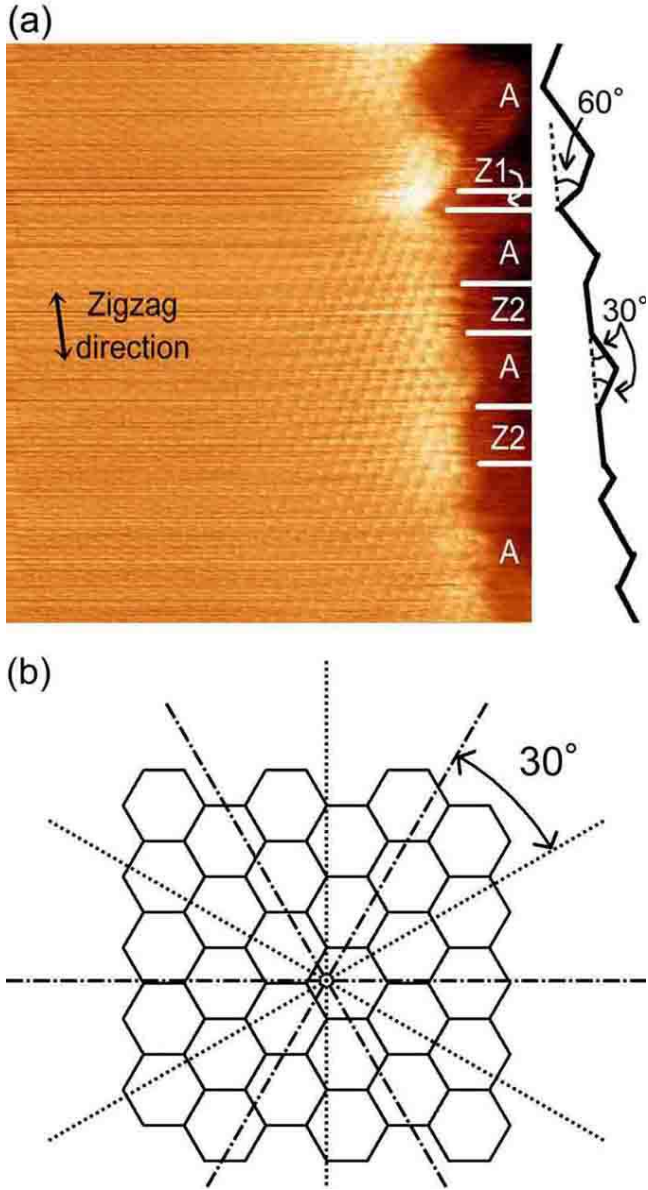}
\caption{\label{fig:edge2}(Color online) (a) STM image (14.8 $\times$ 13.4 nm$^{2}$) of a graphite edge that runs in the zigzag direction. The edge consists of a combination of zigzag 
and armchair edges. The labels `Z1' and `Z2' denote partial zigzag edges that run in different directions, and the label `A' denotes a partial armchair edge. The contour of 
the edge is drawn in parallel to the edge for clarifying positions and angles of the partial zigzag and armchair edges. (b) Schematic model of zigzag and armchair edges that 
are made by cutting along dotted and dash-dotted lines, respectively. Zigzag and armchair edges alternately appear by rotating a cut line by every 30$^{\circ}$.}
\end{figure}

In contrast to the long armchair edge as shown in Fig. 1(a), zigzag edges were typically short. Figure 2(a) shows a typical image observed at an edge whose mean direction 
runs along a zigzag direction of the top layer of graphite. A schematic model of relation of the zigzag/armchair edge and their directions are drawn in Fig. 2(b). By applying 
the relation to Fig. 2(a), it turns out that the proportion of a zigzag edge is less than that of an armchair edge. The lengths of zigzag edges were usually less than several nanometers. 
In Fig. 2(a), there are two types of partial zigzag edges termed as `Z1' and `Z2', whose directions are tilted by 60$^{\circ}$ and 0$^{\circ}$ with respect to the mean edge direction, 
respectively. There are some differences in the observed distributions of bright spots between the `Z1' and `Z2' zigzag edges. The edge carbon atoms of the `Z1' edge are 
brighter than those of the `Z2' edge. In addition to this, the bright spots of the `Z1' edge are limited to the vicinity of the edge carbon atoms and their neighboring sites, 
whereas the bright spots of the `Z2' edge are extended to the interior of the plane.

\begin{figure}[here]
\includegraphics[width=8.5cm, clip]{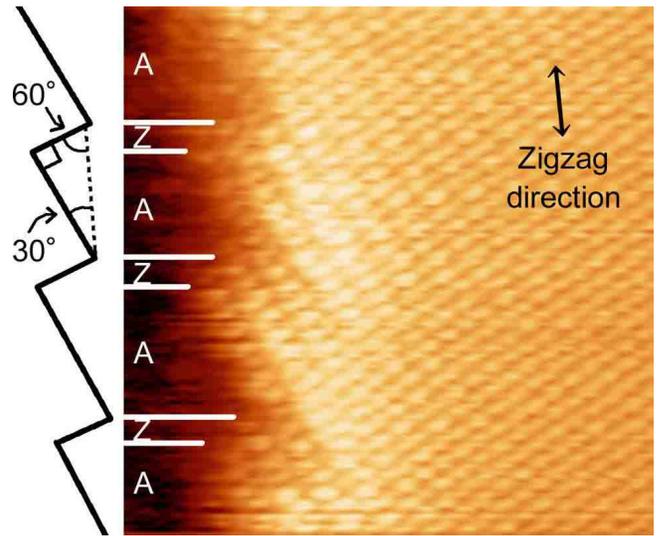}
\caption{\label{fig:edge3}(Color online) STM image (6.7 $\times$ 6.7 nm$^{2}$) of alternate zigzag and armchair edges that run in the zigzag direction. Labels `Z' and `A' denote a partial 
zigzag edge and a partial armchair edge, respectively. The contour of the edge is drawn in parallel to the edge for clarifying the positions and angles of the partial zigzag 
and armchair edges.}
\end{figure}

Another type of an alternate zigzag and armchair edges that run in the zigzag direction is shown in Fig. 3. Brighter spots of the larger LDOS were extended to the 
interior of the plane in Fig. 3. It stands in contrast with large localized LDOS observed on a zigzag edge in Fig. 2(a). It is probably due to different local geometric 
structures that have strong influence on LDOS. Details will be discussed later.

\begin{figure}[here]
\includegraphics[width=8.5cm, clip]{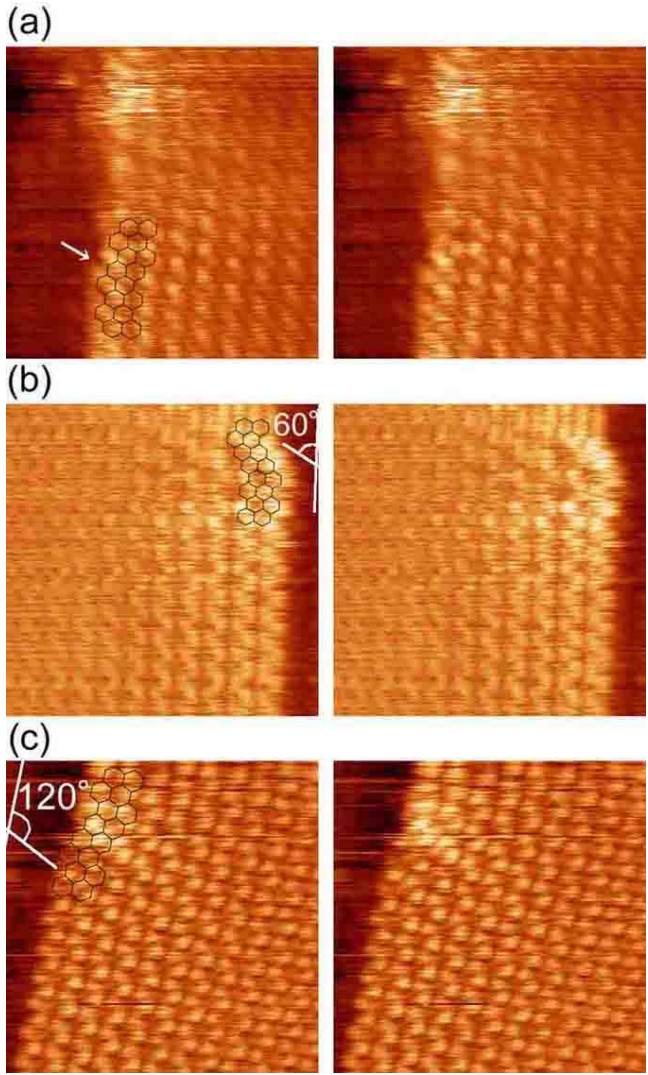}
\caption{\label{fig:edge4}(Color online) (a) STM image (4.0 $\times$ 4.0 nm$^{2}$) of an armchair edge with a defect from which one row of carbon atoms is added to the armchair edge 
in the lower region of the edge. (b, c) STM images ((b) 4.3 $\times$ 4.3 nm$^{2}$, (c) 4.5 $\times$ 4.5 nm$^{2}$) of armchair edges with a defect from which two rows of carbon atoms are added to the lower 
region of the armchair edge. In the left panels, honeycomb lattices and an arrow are overlaid on the original image in the right panels for clarifying the location of the defects. 
Arrays of bright spots were observed at defect points in (b) and (c), and their directions and the angles measured from the direction of the armchair edges are drawn in the left panels.}
\end{figure}

Figure 4 shows STM images of armchair edges with defects from which one (for (a)) or two rows (for (b) and (c)) of extra carbon atoms are attached to armchair edges, 
respectively, judged by applying the honeycomb lattices. Brighter spots were observed around the defects in each image, but their distributions were different depending 
on the local geometric structures.

In contrast to the dispersed LDOS near the defect point in Fig. 4(a), two arrays of bright spots, whose brightness decreased monotonically toward the interior of the 
plane along a line with an angle of 60$^{\circ}$ from the direction of the armchair edge and along the lower armchair edge, were observed near the defect point in Fig. 4(b). 
Similar to the LDOS near the defect point in Fig. 4(b), an array of bright spots, whose brightness decreases monotonically toward the interior of the plane along a line 
with an angle of 120$^{\circ}$ from the direction of the armchair edge, was observed near the defect point in Fig. 4(c).

\begin{figure}[here]
\includegraphics[width=8.5cm, clip]{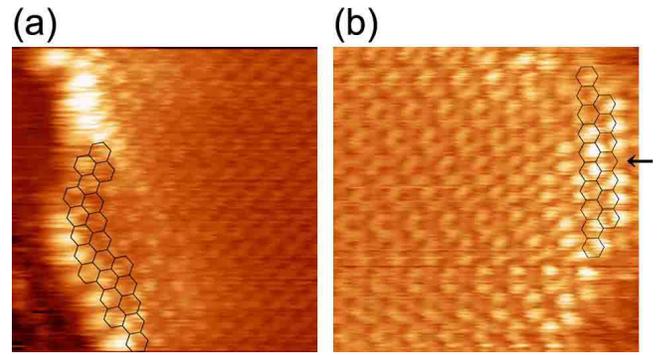}
\caption{\label{fig:edge5}(Color online) (a) STM image (4.5 $\times$ 4.5 nm$^{2}$) of a staircase of partial zigzag edges that consist of five and two edge carbon atoms. (b) STM image 
(3.8 $\times$ 3.8 nm$^{2}$) of a long zigzag edge that consists of seven edge carbon atoms. The honeycomb lattice is drawn for clarifying the location of the zigzag edge. The arrow 
indicates a very small LDOS at the center part of the zigzag edge in (b).}
\end{figure}

Figure 5(a) shows a staircase of a zigzag edge. Bright spots were observed at edge carbon atoms of the partial zigzag edges, and their brightness of the edge carbon 
atoms was not uniform among the edge carbon atoms in each zigzag line that was constituted by two and five carbon atoms. Figure 5(b) shows an STM image of a 
zigzag edge whose length is larger than that of typically observed zigzag edges. The length of the partial zigzag edge consists of seven edge carbon atoms, as can be 
seen by comparing the image with the honeycomb lattice. It should be noted that a kind of node was observed at the point indicated by an arrow.

\section{\label{sec:level4}Discussion}

The experimental results prove that zigzag edges are much smaller in length than those of armchair edges and less frequently observed. Because carbon atoms removed 
in the heat treatment of the sample preparation process are limited to atoms at edges, the ratio of zigzag edges to armchair edges is considered to be conserved in the pristine state. 
Therefore, an armchair edge must be formed more easily than a zigzag edge in its cleavage. This suggests that the structure of an armchair edge is energetically more 
stable than that of a zigzag edge. This agrees with calculated results of carbon nanotubes and nanographene ribbons that have zigzag and armchair edges 
\cite{tomanek2,kawai}. The stability of an armchair edge is higher than that of a zigzag edge in terms of the total energy.

Bright spots are emerged near graphite edges near the Fermi energy ($V_{S} = 0.02$ V) in the STM observations. These spots are not attributed to elevated heights at the 
graphite edges since every layer of graphite does not have any deformation from its flat sheet structure. Instead, at the sample bias, these spots originate from the LDOS, 
which are assigned to edge states near the Fermi energy ($V_{S} = -0.03$ V for a peak of the LDOS) mainly, as shown in STS measurements of the previous report 
\cite{ykobaya}. According to the report, these spots are observed not only at a zigzag edge but also at defects in an armchair edge. In the present paper, further information 
on the distributions of the observed bright spots is given in terms of the LDOS dependent on its local geometric structure.

\begin{figure}[here]
\includegraphics[width=8.5cm, clip]{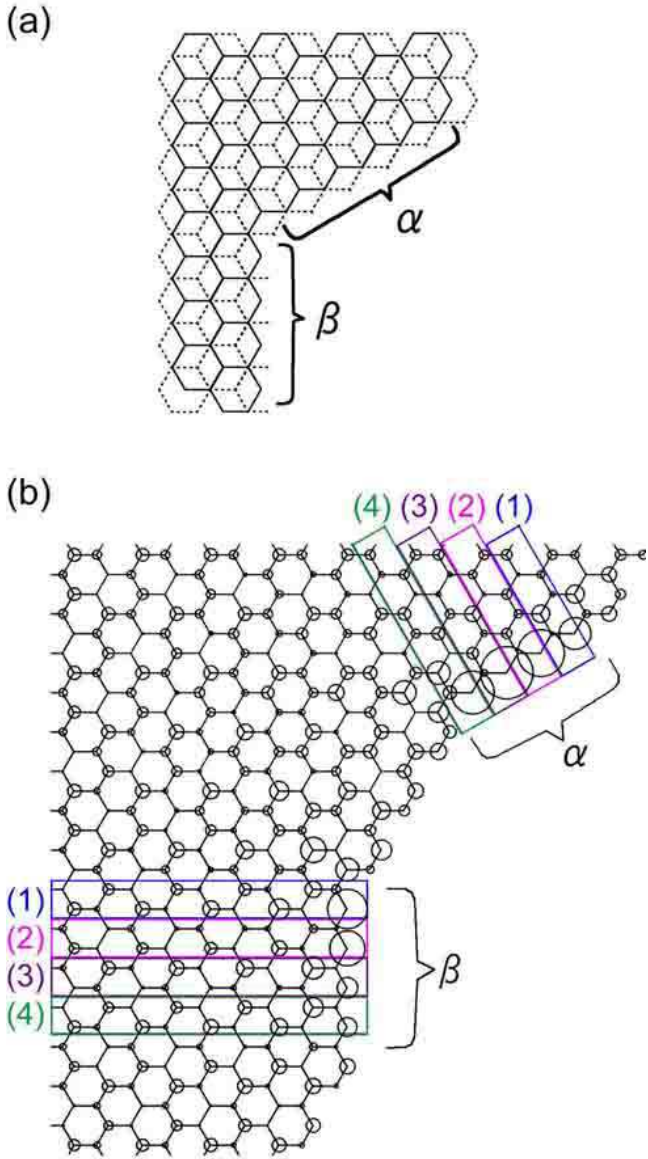}
\caption{\label{fig:edge6}(Color online) (a) Schematic model of $AB$-stacked double-layer graphene that forms two types of zigzag edges whose edge carbon atoms belong to 
$\alpha$-/$\beta$-sites. The angle between the two types of zigzag edges is 60$^{\circ}$. Honeycomb lattices drawn by solid and dotted lines represent the top and second 
layers, respectively. (b) 2D mapping of the LDOS of partial zigzag edges whose directions differ by 60$^{\circ}$ from each other. They are separated with an armchair to 
avoid interference between the edge states of two partial zigzag edges. Each zigzag edge in the model consists of four edge carbon atoms. The $\alpha$-site ($\beta$-site) 
edge carbon atoms of the partial zigzag edges exist at the top right (bottom) part. The LDOS of $\alpha$-site ($\beta$-site) carbon atoms in the rectangles are plotted in Fig. 7.}
\end{figure}
\begin{figure}[here]
\includegraphics[width=8.0cm, clip]{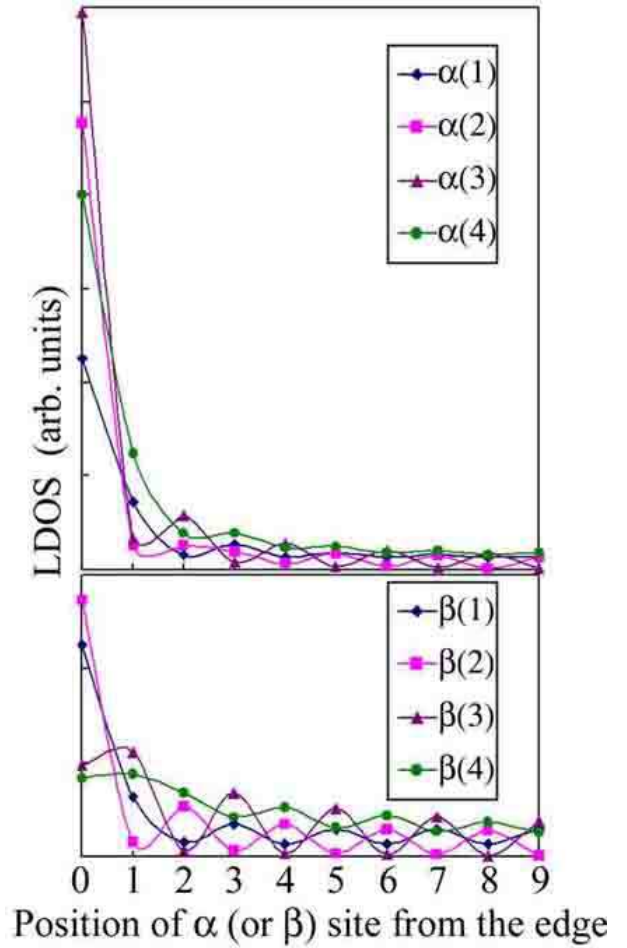}
\caption{\label{fig:edge7}(Color online) Plots of the LDOS in the rectangles numbered in Fig. 6(b) as a function of the position of $\alpha$-site ($\beta$-site) carbon atoms from the 
$\alpha$-site ($\beta$-site) edge carbon atoms. Solid lines are the guides for the eyes.}
\end{figure}

First of all, discussion is devoted to the electronic structure of partial zigzag edges in the edge whose mean direction runs along a zigzag direction, as shown in Fig. 2. 
As noted above, the bright spots of the `Z1' edge are limited to the vicinity of the edge carbon atoms and their neighboring sites, whereas the bright spots of the `Z2' 
edge are extended to the interior of the plane. The origin of the difference in the distribution of the observed bright spots can be understood by a schematic model 
in Fig. 6(a), in which the stacking nature of graphite is taken into account, and two types of zigzag edges are connected with each other with an angle of 60$^{\circ}$. 
The geometric relation in Fig. 6(a) is constructed to extract the relation of the `Z1' and `Z2' edges in Fig. 2(a). The relation in Fig. 6(a) indicates that the edge carbon 
atoms at the top right and the bottom correspond to $\alpha$- and $\beta$-site (or $\beta$- and $\alpha$-site) carbon atoms, respectively. For the calculation of the LDOS 
to be compared with the experimental result in Fig. 2(a), a model is constructed as shown in Fig. 6(b). In this model, the two types of zigzag edges have the same 
geometrical relation to that in Fig. 6(a), except for the presence of the armchair edge that decreases the overlap of the edge states between the two types of zigzag edges. 
In Fig. 7, the LDOS at each $\alpha$-site ($\beta$-site) carbon atoms in the rectangles in Fig. 6(b) is plotted as a function of the position of the $\alpha$-site ($\beta$-site) 
carbon atom from the $\alpha$-site ($\beta$-site) edge carbon atom. The LDOS shows a decay from the edge to the interior with an oscillation that comes from a superperiodic 
pattern near the edges. Figure 7 shows a clear difference in the LDOS distributionsat the partial zigzag edges terminated by the $\alpha$- and $\beta$-site carbon atoms.
The magnitudes of the LDOS of the edge carbon atoms at $\alpha$-sites are larger than those at $\beta$-sites.
The LDOS whose edge carbon atoms are assigned to the $\alpha$-site decays rapidly to the interior of the plane, whereas that at $\beta$-sites decays slowly. In other words, 
the LDOS is more localized at the zigzag edge of $\alpha$-sites, in contrast to the extended feature in that of $\beta$-sites. These results qualitatively reproduce the distribution of the 
bright spots in Fig. 2(a). With this excellent agreement of the characters between the observed and calculated results, the edge carbon atoms of the `Z1' and 
the `Z2' edges correspond to the $\alpha$- and $\beta$-site carbon atoms, respectively. It should be noted that other explanations might be possible for the difference in the 
distributions of the bright spots in Fig. 2(a), because there is little information on the detailed structure of the edges in the STM observation. Klein edges can be created 
instead of zigzag edges after the heat treatment and the subsequent hydrogen-termination in the sample preparation process, and Klein and zigzag edges cannot be 
distinguished by the present STM results. The presence of a Klein edge embedded in a partial zigzag edge can also modify the LDOS distribution \cite{klein1,klein2}.

\begin{figure}[here]
\includegraphics[width=8.5cm, clip]{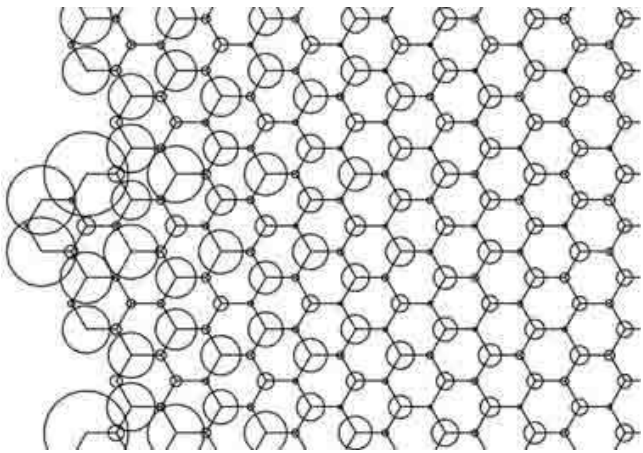}
\caption{\label{fig:edge8} 2D mapping of the LDOS at an edge in the zigzag direction, whose local structure is a simple model to represent that in Fig. 3.}
\end{figure}

The bright spots extended from the alternate zigzag/armchair edges into the interior of the plane in Fig. 3 are interesting, because the slow decay of the LDOS observed 
at the edges is not consistent with the theoretical suggestion on the distribution of edge state at a homogeneous zigzag or armchair edge. The LDOS of the alternate 
zigzag/armchair edges in Fig. 3 is reproduced in the calculated result shown in Fig. 8. From the calculated results, it is obvious that the edge state can be spread widely 
into the interior of the plane. This is because an armchair edge that is sandwiched by partial zigzag edges is smaller in length than the decay length of the edge state of 
the partial zigzag edges. Therefore the mixing of the electronic state at the armchair edges with the edge state, which is extended from the zigzag edges, brings about the 
behavior of a slow decay in the LDOS. If the armchair edges bridging zigzag edges of nanographene are smaller than the decay length, the edge state can be extended 
over its plane, in spite of the tendency of the localization of the edge state at a zigzag edge of a semi-infinite system.

Next, we discuss about the electronic structure of defects in an armchair edge for surveying the origin of the edge state. A small periodic pattern can be seen near the 
armchair edge in Fig. 1(a). Each point of the pattern does not correspond to the usually observed triangle lattice for graphite, but to the superlattice whose periodicity is 
described as $(\sqrt{3}\times\sqrt{3})R30^\circ$. The pattern is considered to originate from the interference waves that are constituted by the incidence wave and the reflected 
wave on graphite surface. The pattern is extended $\sim$10 nm from the edge to the interior of the plane, and its distribution is independent of the presence of defects, 
which may have the edge state. Therefore the presence of the edge states gives no serious effect on the superlattice.

Among defects in armchair edges, the simplest is considered to be dent. Figure 1(b) is a kind of dent, from which basic information on a difference from a homogeneous 
armchair edge (or the effect of edge state) can be obtained. By applying the honeycomb lattice to the observed image, partial zigzag edges are recognized in the structure 
of the dent. To obtain the information about the distribution of the edge state at the dent, calculations of the electronic structure were carried out on the basis of the 
atomic arrangements that were obtained from the experimental result. The 2D LDOS mapping of the calculated model that reproduces the experimental result in Figs. 1(b) 
is shown in Fig. 9.

\begin{figure}[here]
\includegraphics[width=5cm, clip]{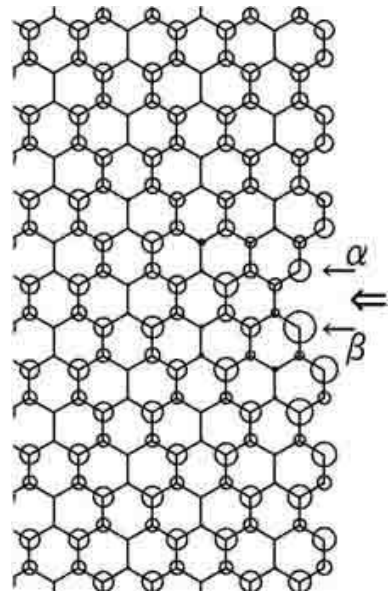}
\caption{\label{fig:edge9} 2D mapping of the LDOS at a defect of the armchair edge, which corresponds to the image of Fig. 1. A thick arrow indicates a dent in the armchair 
edge. Thin arrows indicate $\alpha$-/$\beta$-site of edge carbon atoms of the dent.}
\end{figure}

The LDOS near the dent in Fig. 9 comes from the two partial zigzag edges that are formed by a removal of two carbon atoms from the armchair edge. Due to the presence 
of the second layer of graphene that is put in $AB$-stacking structure, the edge carbon atoms of the two partial zigzag edges have different environments. The edge 
carbon atoms that were bonded to the lost two carbon atoms of the dent belong to $\alpha$- and $\beta$-sites as shown in Fig. 9. Interestingly, the magnitude of 
the LDOS is different between the $\alpha$- and $\beta$-site edge carbon atoms at the partial zigzag edges in the dent. The larger LDOS of the dent in Fig. 9 is located at the 
$\beta$-site atom at one partial zigzag edge, whereas $\alpha$-site atom of another partial zigzag edge of the dent has smaller LDOS. The same difference appears in the 
observed image in Fig. 1(b). Accordingly, the site selectivity helps us to determine the way of stacking and a relative position of the second layer, because models of 
mono-layered graphene and of $AA$-stacked graphite cannot reproduce the site selectivity.

\begin{figure}[here]
\includegraphics[width=8.5cm, clip]{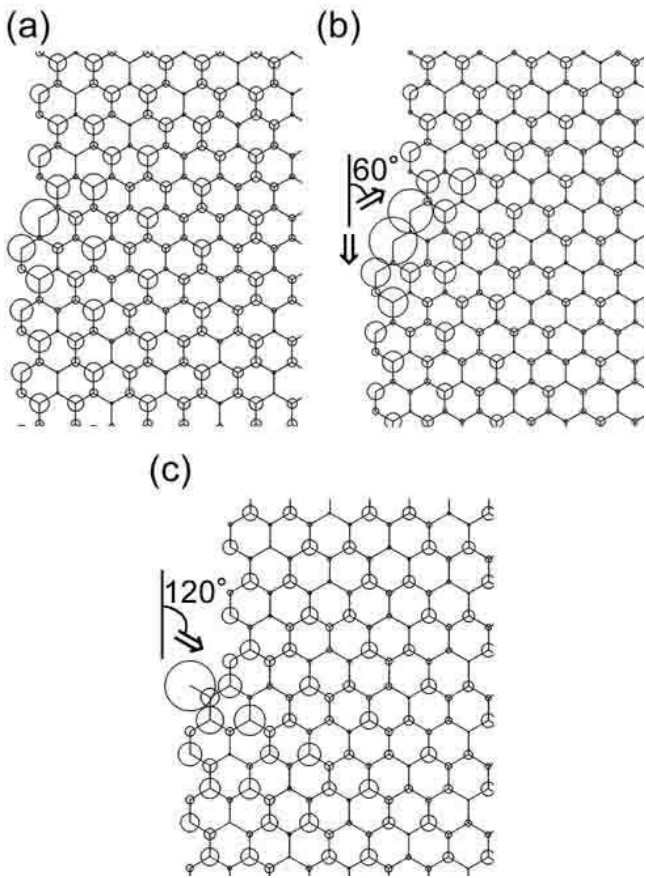}
\caption{\label{fig:edge10} 2D mappings of the LDOS at defect points of the armchair edges. (a), (b), and (c) correspond to the images of Figs. 4(a), 4(b), and 4(c), respectively. 
In (c), an extra carbon atom, which is drawn as a bar, is attached to an edge carbon atom of the partial zigzag edge. Arrows indicate the direction of the arrays of large LDOS 
that is extended to the interior of the plane.}
\end{figure}

As the third issue, we discuss a subject how the character of a partial zigzag edge embedded in an armchair edge is changed if the number of edge carbon atoms of the 
partial zigzag edge increases as shown in Fig. 4. Figures 4(a) and 4(b) show partial zigzag edges embedded in an armchair edge, where one and two edge carbon atoms 
at the partial zigzag edges are present, respectively. In contrast with a dispersed LDOS distributed near the defect point in Fig. 4(a), two arrays of bright spots appear 
as shown in Fig. 4(b) by increasing just one edge carbon atom at the partial zigzag edges. The observed difference between Figs. 4(a) and 4(b) is reproducible in the 
calculated results in Figs. 10(a) and 10(b). The arrays of bright spots in Fig. 4(b) are reproduced as arrays of the LDOS along a line with an angle of 60$^{\circ}$ from the 
direction of the armchair edge and along the lower armchair edge in Fig. 10(b). Therefore the difference in the distributions of the LDOS originates from the difference in the 
number of edge carbon atoms involved in the partial zigzag edges.

Figure 4(c) is the STM image of an edge whose structure has a defect similar to that of Fig. 4(b) in a sense that the two rows of carbon atoms are attached to an armchair edge. 
However, the direction of the array of bright spots in Fig. 4(c) is different from that in Fig. 4(b). The array of bright spots in Fig. 4(c) runs with their brightness 
decreasing toward the interior of the plane along a line with an angle of 120$^{\circ}$ from the direction of the armchair (if the angle is measured in the same way 
with that of Fig. 4(b)). Here, we remind the difference in the stacking structures, which is an excellent explanation for the case of Fig. 2(a). However, the difference between 
the stacking manners, that is, the difference of $\alpha$-/$\beta$-site at the defect point or the difference between $AA$- and $AB$-stacking, is not a candidate for the origin 
of the difference between the two STM images, because the observed difference cannot be reproduced on the basis of the calculations with the difference of 
$\alpha$-/$\beta$-site. The difference that originates from the local stacking structure can create only the difference of the magnitude of LDOS like in the case of 
Fig. 6(b) and do not give any difference in the direction of an array of relatively large LDOS. Hence, the origin of the difference in the direction of the array of bright 
spots may come from the difference in detailed edge carbon structure that cannot be clearly distinguished by STM. The origin of the difference may be associated 
with the presence of an extra carbon atom that is bonded to the edge after the sample preparation process, for example. The smallest structural difference 
at the defect point can be made by the presence/absence of one carbon atom attached to the edge carbon atom, that is, presence/absence of a Klein edge at the 
defect. Figure 10(c) is the calculated result based on the model that has the same geometric structure as that of Fig. 10(b) except that an extra carbon atom is attached 
to the partial zigzag edge. Figure 10(c) well reproduces the array of bright spots in Fig. 4(c). The direction of an array of the LDOS of Fig. 10(c) is 120$^{\circ}$ from the 
direction of the armchair edge and is changed by 60$^{\circ}$ from that of Fig. 10(b) by the presence of a Klein edge.

\begin{figure}[here]
\includegraphics[width=8.5cm, clip]{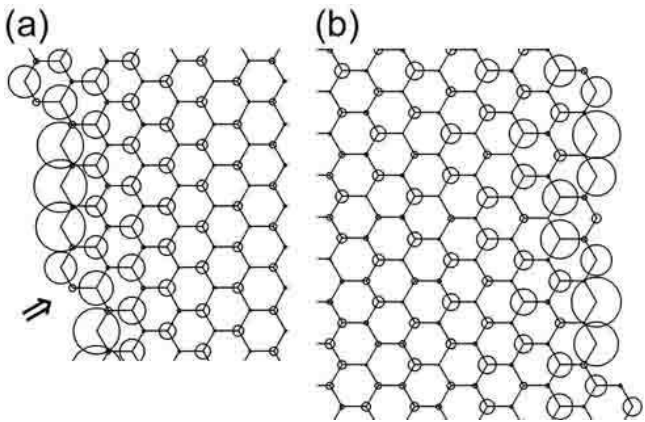}
\caption{\label{fig:edge11} 2D mappings of the LDOS at relatively long zigzag edges. Local geometric structures of (a) and (b) are simple models to represent Fig. 5(a) and 5(b), 
respectively. An arrow in (a) indicates a defect point that corresponds to a partial armchair edge.}
\end{figure}

Finally, we discuss the electronic structure of a relatively long zigzag edge embedded in an armchair edge. The staircase-like zigzag edge that consists of the partial zigzag 
edges is one of the relatively long zigzag edges observed (Fig. 5(a)). It revealed that the edge carbon atoms at the zigzag edge in Fig. 5(a) are not necessarily observed 
as bright spots. For clarifying the origin of the nonuniformly distributed bright spots, LDOS calculation was carried out for the model whose unit cell consisted of 
four edge carbon atoms of the zigzag edge as a simplified model to obtain basic features of the staircase-like zigzag edge in Fig. 5(a). In Fig. 11(a), the LDOS that is 
varied among edge carbon atoms of the partial zigzag edge comes from the electron confinement effect in the partial zigzag edge. The LDOS at the defect point of the 
partial zigzag edge, which is indicated by an arrow in the figure, is smaller than that of other carbon atoms of the edge since the defect point of the partial zigzag edge 
corresponds to the node of the wave function.

Another type of relatively long zigzag edge in Fig. 5(b) consists of seven edge carbon atoms. In this image, the bright spot that represents the edge state is absent 
at the center of the zigzag edge. The absence of bright spots was always detected in a partial zigzag edge that is longer than that in Fig. 5(b). In the calculated result 
in Fig. 11(b), the LDOS at the center part of the partial zigzag edge is very small although the atomic structure does not have any vacancy in the edges. This is because 
the smallest LDOS in the edges corresponds to the node of wave function of an electron that is confined at the partial zigzag edge.

\section{\label{sec:level5}Conclusion}

Edge states of graphite, which are strongly dependent on local edge structures, have been investigated by STM for hydrogen-terminated step edges. Armchair 
edges are formed at the steps of graphite more frequently than zigzag edges, and their length tends to continue in hundreds nm. In contrast to the long armchair 
edge, zigzag edges are short. The LDOS associated with the edge state at the zigzag edges have different distributions dependent on the $\alpha$- or $\beta$-site of edge carbon atoms at the zigzag 
edge. Edge states at defect points of armchair edges are extended to a neighbored armchair edge and the interior of the plane, dependent on the detailed edge 
structure constituted by partial zigzag/Klein edge. The origin of the inhomogeneous feature in the LDOS distribution of longer zigzag edges is attributed to the 
electron confinement effect in the partial zigzag edge.

Detailed information on atomic arrangement of edges that are the basis of calculations is required experimentally in the future work for understanding the correlation 
edge states dependent on detailed edge structures. Knowledge accumulated from the information is expected to contribute to material design of the electronic device based on graphite edge.
\begin{acknowledgments}
The authors thank A. Moore for his generous gift of the HOPG substrate. The present work was supported by Grant-in-Aid for Scientific Research No. 15105005 
from the Ministry of Education, Culture, Sports, Science, and Technology, Japan.
\end{acknowledgments}
\bibliography{apssamp}
\end{document}